# Topological quantum phase transition and superconductivity induced by pressure in the bismuth tellurohalide BiTeI

Yanpeng Qi[1], Wujun Shi[1,2], Pavel G. Naumov[1,3], Nitesh Kumar[1], Raman Sankar[4,5], Walter Schnelle[1], Chandra Shekhar[1], F. C. Chou[5], Claudia Felser[1], Binghai Yan[1,2,6]*, Sergey A. Medvedev[1]*

[1]Max Planck Institute for Chemical Physics of Solids, 01187 Dresden, Germany.
[2]School of Physical Science and Technology, ShanghaiTech University, Shanghai 200031, China.
[3]Shubnikov Institute of Crystallography of Federal Scientific Research Centre "Crystallography and Photonics" of Russian Academy of Sciences, 119333, Moscow, Russia
[4]Institute of Physics, Academia Sinica, Taipei 10617, Taiwan.
[5]Center for Condensed Matter Sciences, National Taiwan University, Taipei 10617, Taiwan.
[6]Max Planck Institute for the Physics of Complex Systems, 01187 Dresden, Germany.

**Abstract**

**A pressure-induced topological quantum phase transition has been theoretically predicted for the semiconductor BiTeI with giant Rashba spin splitting. In this work, the evolution of the electrical transport properties in BiTeI and BiTeBr is investigated under high pressure. The pressure-dependent resistivity in a wide temperature range passes through a minimum at around 3 GPa, indicating the predicted transition in BiTeI. Superconductivity is observed in both BiTeI and BiTeBr while the resistivity at higher temperatures still exhibits semiconducting behavior. Theoretical calculations suggest that the superconductivity may develop from the multi-valley semiconductor phase. The superconducting transition**

**temperature $T_c$ increases with applied pressure and reaches a maximum value of 5.2 K at 23.5 GPa for BiTeI (4.8 K at 31.7 GPa for BiTeBr), followed by a slow decrease. Our results demonstrate that BiTe$X$ ($X$ = I, Br) compounds with non-trivial topology of electronic states display new ground states upon compression.**

___________________________________

[*] E-mail: Sergiy.Medvediev@cpfs.mpg.de; Yan@cpfs.mpg.de

**Introduction**

The physics of the layered bismuth tellurohalides (BiTe$X$ with $X$ = I, Br and Cl) with giant Rashba spin splitting, which may facilitate future spintronic applications, has found widespread interest[1,2]. Recently, first-principles band-structure calculations have suggested that BiTeI undergoes a transition to a topological insulating phase under pressure ($P$)[3]. Unlike the previously discovered topological insulators with inversion symmetry, the inversion symmetry in bismuth tellurohalides is naturally broken in the crystal structure[4]; therefore, the realization of topological magneto-electric effects[5] and topological superconductivity within the Majorana edge channels[6] is highly possible.

However, experimental verification of predicted phenomena such as the topological quantum phase transition (TQPT) remains highly controversial. For example, infrared spectroscopy investigations on BiTeI have drawn contradictory conclusions, i.e., the presence[7] or absence[8] of the TQPT at high pressure. Recent studies of pressure-dependent electrical transport have proved insufficient to discover the TQPT due to a range that possibly did not comprise the critical pressure $P_c$ [9,10,11]. Up to now there is no consistent picture for the transport properties in the expected critical range of pressure in BiTeI or BiTeBr. On the other hand, a pressure-induced structural phase transition $P3m1 \rightarrow Pnma$ was already reported for BiTeI[12]. Therefore the transport properties of BiTeI and BiTeBr under pressure are of current research interest.

In this work, we systematically investigate the high-pressure behavior of BiTeI as well as of the sister compound BiTeBr (most details for BiTeBr are documented in the Supplementary Information). Through electrical transport measurements, we find that the application of pressure initially leads to a slight suppression of the overall magnitude of resistivity, followed by an enhancement resulting in a minimum at around 3 GPa, indicating a TQPT in BiTeI. At certain pressures higher than the transition pressure to the *Pnma* phase superconductivity is observed in both BiTeI and BiTeBr, even while the normal state continues to exhibit non-metallic behavior. Superconductivity persists in the *P*4/*nmm* phases of both compounds at least up to $P \approx 40$ GPa. Maximum critical temperatures $T_c$ of 5.2 K at 23.5 GPa and of 4.8 K at 31.7 GPa are observed for BiTeI and BiTeBr, respectively.

**Results**

**Structure and transport properties under ambient pressure.** The BiTe*X* (*X* = I or Br) single crystals used in this study were grown via a chemical vapor transport method. BiTe*X* compounds crystallize in the non-centrosymmetric space group of *P*3*m*1 (Fig. 1a). In the triple layer structure stacked along the *c* axis the Bi atom layers are sandwiched between one Te and one halogen layer. Despite the strong chemical bonding within each such unit, the adjacent triple layers are weakly coupled via the van der Waals interaction.

Although electronic structure calculations indicate that both BiTe*X* (*X* = I and Br) compounds are semiconductors, the measured temperature (*T*) dependence of the electrical resistivity, ρ(*T*), for both BiTeI and BiTeBr (Supplementary Fig. 1a and c) reveals a metallic-like behavior apparently as a result of self-doping due to deviations from stoichiometry. The Hall resistivities $\rho_{yx}$ for both BiTeI and BiTeBr (Supplementary Fig. 1b and d) are linearly dependent on magnetic field in the entire investigated range of magnetic fields and have negative slope at all *T*, indicating that the transport properties of both compounds are governed only by electron-type charge carriers. This finding is consistent with recent angle-resolved photoemission spectroscopy (ARPES) experiments[13] and thermopower measurements[14]. Within a single-band model the electron concentration $n_e$ is estimated to $4.0 \times 10^{19}$ and $0.42 \times 10^{19}$ cm$^{-3}$ at 2 K for BiTeI and BiTeBr, respectively, which is close to previously reported values[10].

**Electrical resistivity at high pressure.** Figure 2 shows the temperature dependence of the resistivity $\rho(T)$ of BiTeI for various pressures. In a low-pressure region, increasing pressure initially induces a weak but continuous suppression of the overall magnitude of $\rho$. Subsequently, resistivity increases, with a minimum occurring at $P_{min}$ = 2.4–3.6 GPa. Upon further increasing the pressure above 3.6 GPa, resistivity starts to increase rapidly reaching a maximum at a pressure above 8 GPa.

With further increasing pressure above 8.8 GPa, $\rho$ gradually decreases for BiTeI (Fig. 2 and Fig. 3a) as well as for BiTeBr at pressures above 7 GPa (Supplementary Fig. 2). As pressure increases up to 10 GPa, a drop of $\rho$ is observed in BiTeI at temperatures below 2 K and zero resistivity is achieved for $P \geq 13.9$ GPa, indicating the emergence of superconductivity. Interestingly, $\rho(T)$ at this pressure still exhibits an non-metallic behavior in the normal state. Similarly, pressure-induced superconductivity with semiconducting normal state is observed in BiTeBr at pressures above 13 GPa (Supplementary Fig. 2) and BiTeCl[15].

The critical temperature ($T_c$) is enhanced with further increase of pressure, and a maximum $T_c$ of 5.2 K is attained at $P$ = 23.5 GPa (Fig. 3a). Beyond this pressure, $T_c$ decreases slowly but superconductivity persists up to the highest experimental pressure of $\approx$ 38 GPa. A similar evolution of $\rho$ is observed for BiTeBr and a maximum $T_c$ of 4.8 K is attained at $P$ = 31.7 GPa (Supplementary Fig. 3a).

Figure 3b displays $\rho(T)$ curves in external magnetic fields for BiTeI at $P$ = 23.5

GPa. The superconducting transition gradually shifts towards lower $T$ with the increase of magnetic field. A magnetic field $\mu_0 H$ = 3 T deletes all signs of superconductivity above 1.8 K. The upper critical field $\mu_0 H_{c2}$ is determined using the 90% points on the $\rho$ transition curves and plots of $H_{c2}(T)$ are shown in Fig. 3c. The slope $d\mu_0 H_{c2}/dT$ at $T_c$ for BiTeI is -0.81 T K$^{-1}$. A simple estimate using the conventional one-band Werthamer-Helfand-Hohenberg (WHH) approximation, but neglecting the Pauli spin-paramagnetism effect and spin-orbit interaction[16], i.e., $\mu_0 H_{c2}(0)$ = -0.693 × $\mu_0(dH_{c2}/dT)$ × $T_c$, yields a value of 2.9 T for BiTeI. We also tried to use the Ginzburg-Landau formula to fit the data:

$$H_{c2}(T) = H_{c2}(0)\frac{1-t^2}{1+t^2} \qquad (1)$$

where $t = T/T_c$. A critical field $\mu_0 H_{c2}$ = 3.0 T results for BiTeI. Both values are well comparable with those determined for superconducting Bi$_2$Se$_3$ under pressure[17, 18]. According to the relationship $\mu_0 H_{c2} = \Phi_0/(2\pi\xi^2)$, where $\Phi_0 = 2.07\times10^{-15}$ Wb is the flux quantum, the coherence length $\xi_{GL}(0)$ is 10.5 nm for BiTeI. It is worth to note that the extrapolated values of $H_{c2}(0)$ are well below the Pauli-Clogston limit. A similar evolution of $\rho$ in external magnetic fields is observed for BiTeBr and shown in Supplementary Fig. 3b,c.

**Discussion**

The pressure dependences of the resistivity in the normal state at $T$ = 6 K and of the critical temperature of superconductivity together with known structural data[12] for BiTeI are summarized in Fig. 4 and for BiTeBr in Supplementary Fig. 5. The presented results demonstrate that high pressure dramatically alters the electronic properties in both BiTeI and BiTeBr. In BiTeI, the resistivity in a wide temperature range first decreases with $P$ to a minimum but then increases strongly. This peculiar behavior of the resistivity of BiTeI is not associated with a structural phase transition since high-pressure x-ray diffraction (XRD) studies revealed the structural stability of the ambient-pressure $P3m1$ structure up to 8.8 GPa[12]. As shown in Supplementary Fig. 6, BiTeI is a normal semiconductor with a narrow band gap ($\Delta E$ = 0.28 eV) at ambient pressure. When $P$ is increased to the critical pressure, $P_c$, the conduction band (CB) minimum and the valence band (VB) top meet at a special $k$ point along the $k_x$ ($A - H$) direction in the Brillouin zone. Under these conditions, the band gap collapses and a zero-gap semimetallic state appears. At higher $P$, the band gap reopens (Supplementary Fig. 6) via inversion of the VB and CB characteristics and the system enters the topological insulator phase. The valley-shaped evolution of electrical resistivity with pressure is consistent with our theoretical calculations and we thus can identify $P_c$ of the TQPT with the pressure of the minimum of the resistivity $P_{min}$. In addition, recent XRD analyses have shown that the lattice parameter ratio $c/a$ of BiTeI passes through a minimum at approximately 3 GPa, indicating an enhanced bonding along the $c$ axis induced by the $p_z$ band crossing[7, 12]. Our results are consistent with these findings and indicate a TQPT in BiTeI at a $P_c \sim$ 3 GPa.

At a further increase of pressure, two structural phase transitions are reported for BiTeI: to the $Pnma$ phase II (Fig. 1b) at 8.8 GPa and to the $P4/nmm$ phase III (Fig. 1c) at 18.9 GPa respectively[12]. Phase II has a unique orthorhombic structure with space group $Pnma$, where one-dimensional zigzag Bi−Te ladders occupy the vertex lattice sites and the I atoms remain in the interstitials. Phase III crystallizes in the $P4/nmm$ space group and forms stacks in a Te−Bi−I−I−Bi−Te sequence along the $c$ axis, which can be characterized as being comprised of interconnected cubic building blocks consisting of Te−Bi−I atoms that form three-dimensional networks. Compared with the *Pnma* structure, the *P4/nmm* structure offers almost perfect atomic arrangement with alternating opposite faces, leading to the densest-possible atomic packing[12]. The enthalpy calculations (Supplementary Fig. 4) demonstrate that BiTeBr also undergoes transitions to the *Pnma* and *P4/nmm* structures.

In order to obtain a comprehensive understanding of the physical properties of the high pressure phases of BiTeI and BiTeBr, we performed density functional theory calculations (DFT) for the electronic band structures. The electronic structures of the *Pnma* phases of both BiTeI and BiTeBr are those of topologically trivial semiconductors. The resistivity is related to the large density of states (DOS) in the CB. For the *Pnma* phase of BiTeI, the CB at the T point of the Brillouin zone shifts towards the Fermi level $E_F$ with increasing pressure (see Fig. 5), which results in an increase of the DOS near $E_F$. Therefore, the resistivity decreases with increasing pressure. The evolution of the gap in BiTeI between the T point in the CB and the X point in the VB is shown in the upper panel of Fig. 4, which is consistent with observed variation of the resistivity with pressure.

The occurrence of superconductivity in the semiconducting *Pnma* phase is reminiscent of the superconductivity proposed for many-valley semiconductors and similar semimetals in the 1960s, which was then experimentally observed in electron-doped $SrTiO_3$[19]. As can be seen in Fig. 5 for BiTeI (Supplementary Fig. 7 for BiTeBr), the band structure exhibits a multi-valley characteristic over the entire stability range of the *Pnma* phase. γ marks two bands at the T point of the Brillouin zone. Here, the effective electron-electron attraction may arise from exchange of the intra-valley and inter-valley phonons, leading to superconductivity[20].

Different from the two former phases, the $P4/nmm$ phase is semimetallic and exhibits a finite DOS at $E_F$ (see Supplementary Fig. 8), which is consistent with our resistivity data. As the superconductivity occurs among electronic states at $E_F$, we investigated the DOS at $E_F$ in the $P4/nmm$ phase for various pressures. It is observed that the DOS decreases monotonically with increasing $P$. This pressure dependence of the DOS agrees well with the variation of $T_c$ shown in the phase diagram (Fig. 4).

**Methods**

**Single–crystal growth and characterization.** BiTeI and BiTeBr crystals were grown via a chemical vapor transport method. The starting materials used for the synthesis of BiTeBr were Bi, Te, $BiBr_3$ and $TeBr_4$. A quartz ampoule was charged with these materials in a ratio corresponding to the stoichiometry of the target compound. After evacuation and sealing, the ampoule was inserted into a furnace with a temperature gradient of 450 to 480 °C, with the educts being in the hot zone. After 100 h, crystals with dimensions of $5 \times 8$ mm$^2$ and thicknesses of several tenths of a millimeter had grown in the cold zone. A similar method was used to grow BiTeI single crystals[21].

**High-pressure experiments.** Resistivity measurements under high pressure were performed in a non-magnetic diamond anvil cell[22, 23]. A cubic BN/epoxy mixture was used for the insulating gaskets and Pt foil was employed for the electrical leads. The diameters of the flat working surface of the diamond anvil and the hole in the gasket were 0.5 and 0.2 mm, respectively. The sample chamber thickness was ≈ 0.04 mm. Resistivity was measured using an inverting dc current in a van der Pauw technique implemented in a customary cryogenic setup (lowest achievable temperature $T_{min}$ = 1.5 K) at zero magnetic field, and the magnetic-field measurements were performed on a magnet-cryostat (PPMS-9, Quantum Design, $T_{min}$ = 1.8 K). Pressure was measured using the ruby scale for small chips of ruby placed in contact with the sample[24].

**Density functional theory calculations.** Density functional theory (DFT) calculations were performed using the Vienna Ab-initio Simulation Package (VASP)[25] with a plane wave basis. The interactions between the valence electrons and ion cores were described using the projector augmented wave method[26, 27]. The exchange and correlation energy was formulated using the generalized gradient approximation with the Perdew-Burke-Ernzerhof scheme[28]. The plane-wave basis cutoff energy was set to 176 eV by default. $\Gamma$-centered $k$ points with a spacing of 0.03 Å$^{-1}$ were used for the first Brillouin-zone sampling. The structures were optimized until the forces on the atoms were less than 5 meV Å$^{-1}$. The pressure was derived by fitting the total energy dependence on the volume using the Murnaghan equation[29]. Note that spin-orbit coupling was included in the static calculation.

**Acknowledgments**

Y. Qi acknowledges financial support from the Alexander von Humboldt Foundation. This work was financially supported by Deutsche Forschungsgemeinschaft (DFG; Project No. EB 518/1-1 of DFG-SPP 1666 "Topological Insulators") and by the European Research Council (ERC Advanced Grant No. (291472) "Idea Heusler").

**Figure Captions**

**Figure 1. Crystal and band structures of BiTeI.** Crystal structures: (**a**) Trigonal $P3m1$ structure at ambient pressure and the high-$P$ phases (**b**) $Pnma$ and (**c**) $P4/nmm$. The blue, red, and green spheres represent Bi, Te, and I atoms, respectively. (**d**)–(**f**) Schematic illustrations of the band structure evolution under various pressures. (**d**) $n$-type semiconductor and topological insulator; (**e**) normal semiconductor; and (**f**) semimetal.

**Figure 2. Evolution of the electrical resistivity as function of pressure for BiTeI.** (**a**) and (**b**) Temperature dependence of the resistivity in the low-pressure range. The resistivity in the full temperature range is suppressed with increasing pressure but increases above the critical pressure $P_c$, indicating the topological quantum phase transition (TQPT). (**c**) The resistivity is gradually suppressed with increasing pressure and superconductivity emerges above ≈ 10 GPa while the resistivity generally exhibits an activated semiconductor behavior. (**d**) Upon heavy compression the resistivity is typically metallic and for $P > 25$ GPa the superconducting transition temperature decreases slowly.

**Figure 3. Temperature-dependent resistivity in the vicinity of the superconducting transition and determination of the upper critical field for BiTeI.** (**a**) Drop of the electrical resistivity and zero-resistance behavior of BiTeI. With increasing pressure, the superconducting critical temperature $T_c$ increases and a maximum $T_c = 5.2$ K is observed at 23.5 GPa. (**b**) Temperature dependence of resistivity under different magnetic fields for BiTeI at 23.5 GPa. (**c**) Temperature dependence of upper critical field for BiTeI. Here, $T_c$ is determined as the 90% drop of the normal state resistivity. The solid lines represent the fits based on the Ginzburg-Landau (GL) formula.

**Figure 4. Electronic *P-T* phase diagram for BiTeI.** The colored areas represent different phases, shaded areas indicate the coexistence of two phases. The upper panel shows the pressure dependence of the calculated band gaps of phase I (gap between T point in conduction band and X point in valence band of phase II) and the density of states (DOS) at the Fermi level for phase III. The lower panel shows the superconducting $T_c$ as function of pressure. The resistivity values at 6 K are also shown. The green and blue circles represent the $T_c$ extracted from different runs of resistivity measurements.

**Figure 5. Calculated Fermi surface and band structure of the *Pnma* phase of BiTeI.** (**a**) Fermi surface. At low pressure two Fermi surfaces appear (α and β). From 10.55 GPa, two new Fermi surfaces (γ) enclose the T point. To plot the Fermi surface, 0.02 electrons per unit cell are doped to align the Fermi energies of the different structures. The red and blue surfaces represent the two bands. (**b**) Band structure. The black dashed and red solid lines represent the Fermi levels without doping and with electron-doping, respectively.

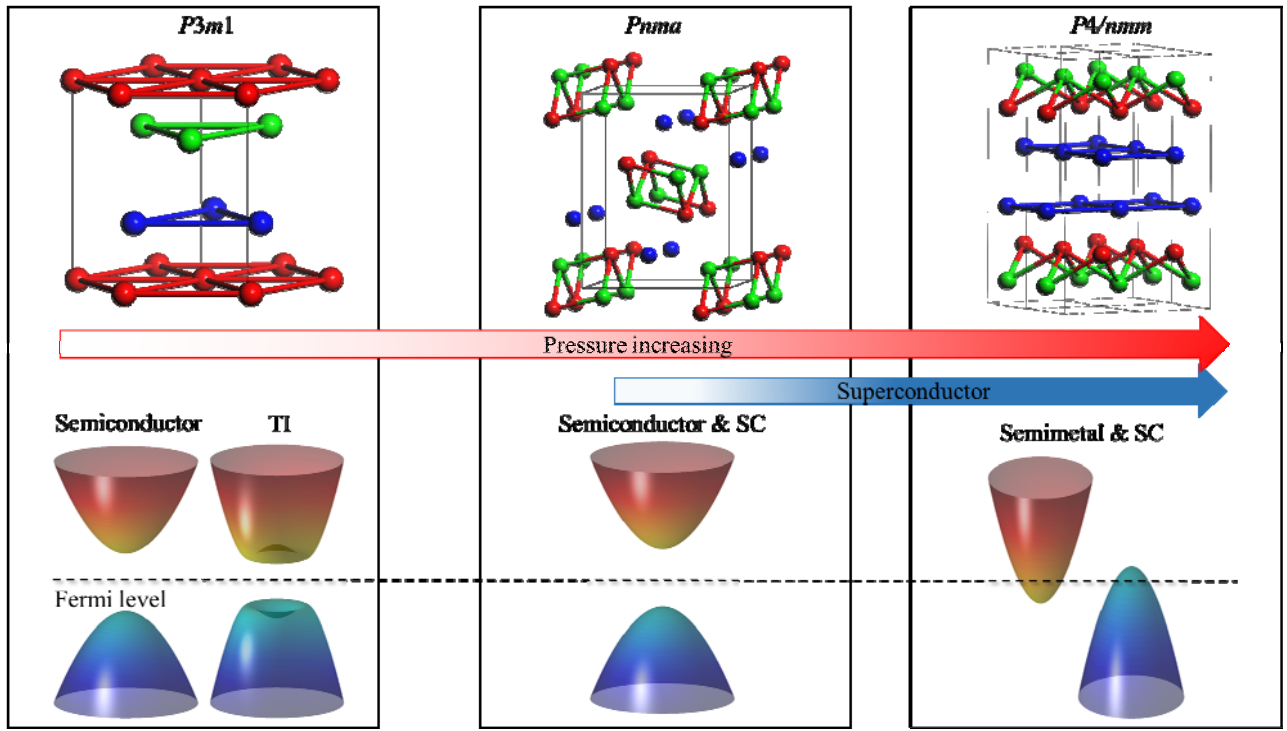

Fig. 1 Qi et al.

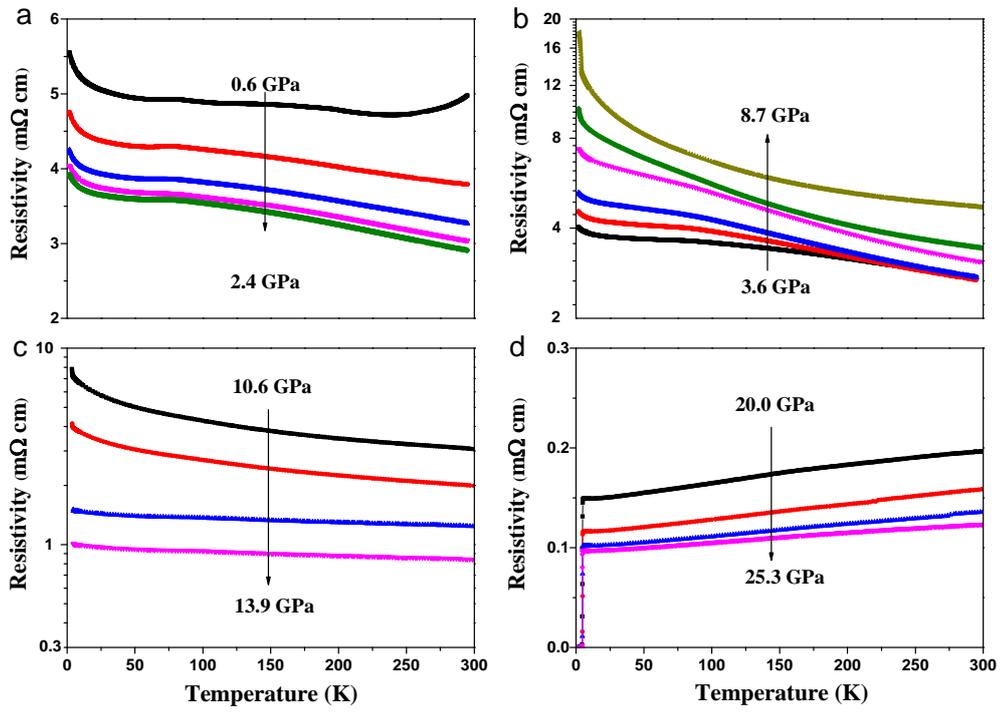

Fig. 2 Qi et al.

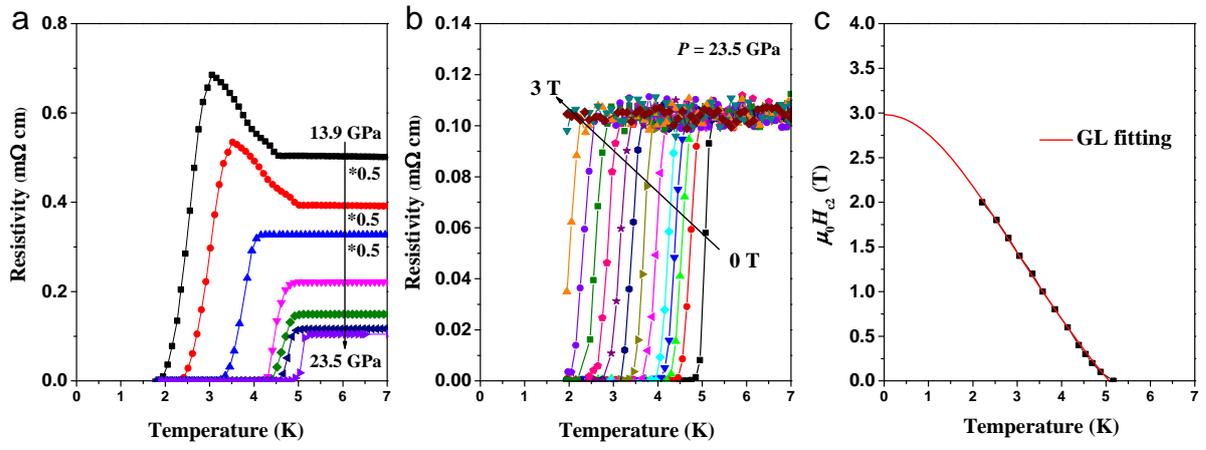

Fig. 3 Qi et al

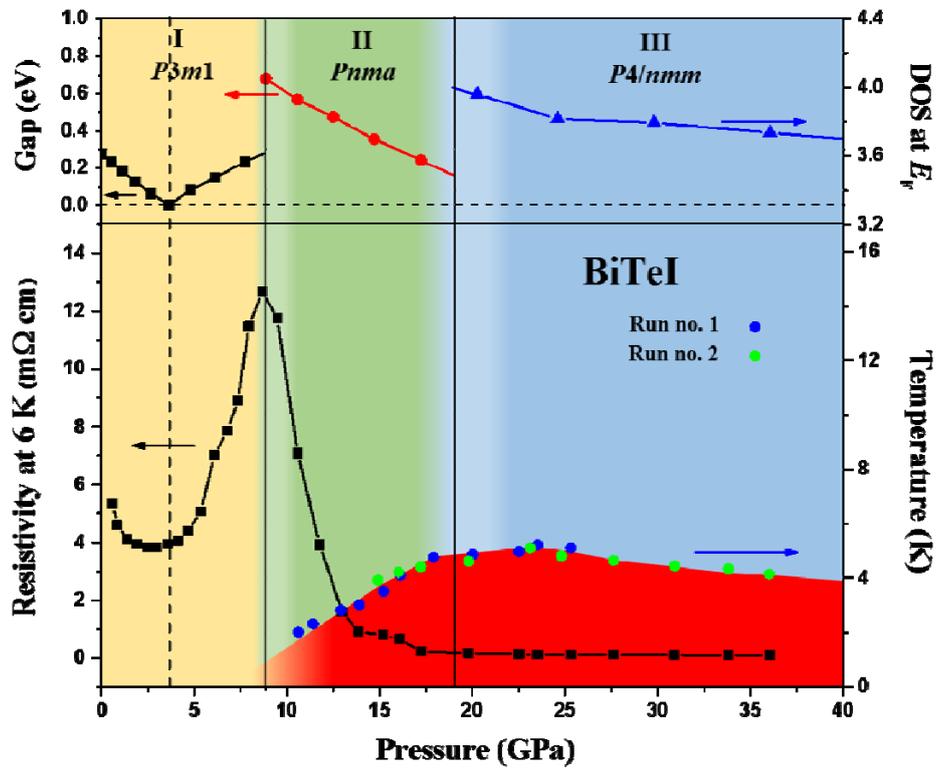

Fig. 4 Qi et al.

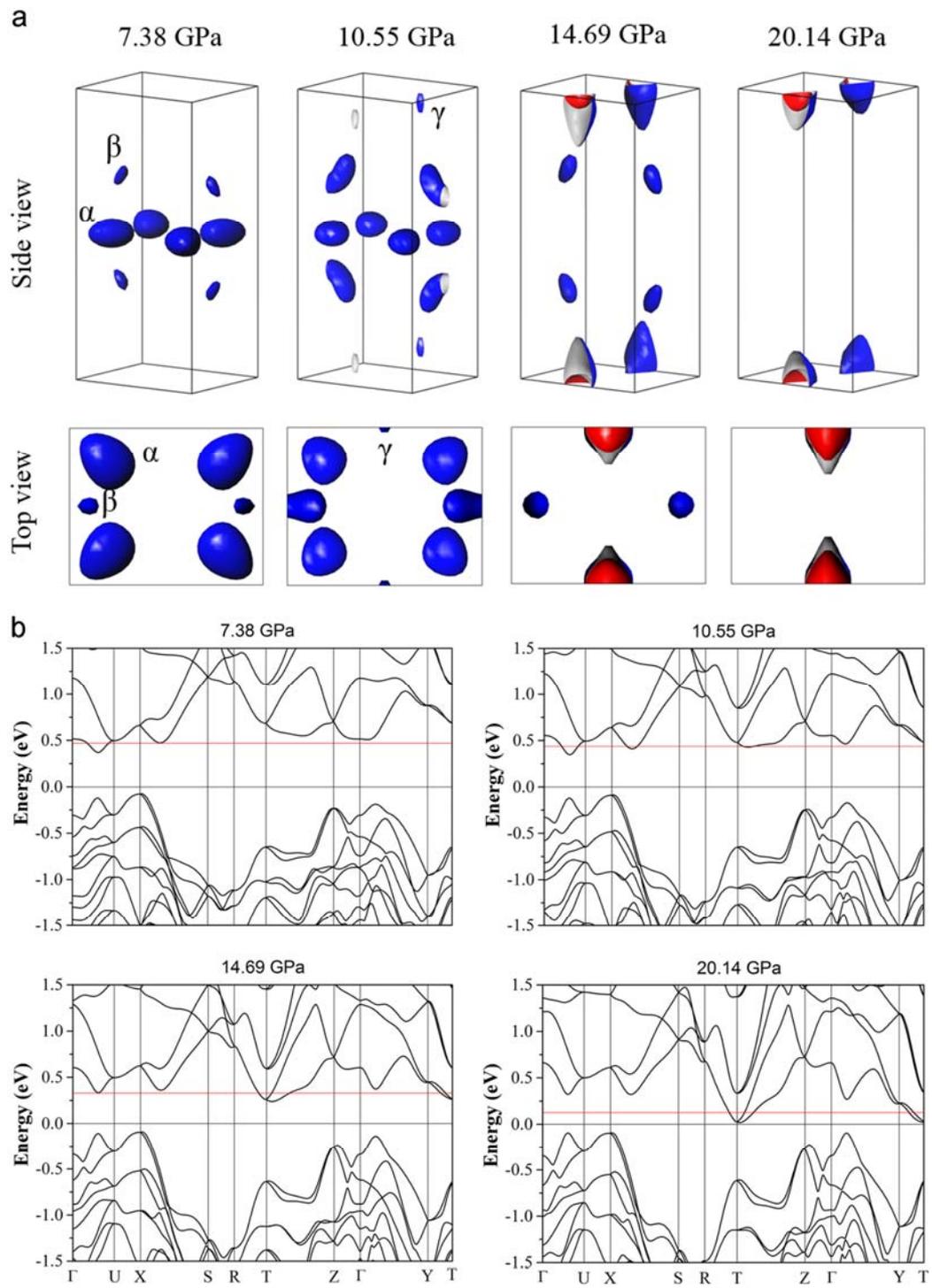

Fig. 5 Qi et al

**Supplementary Figures**

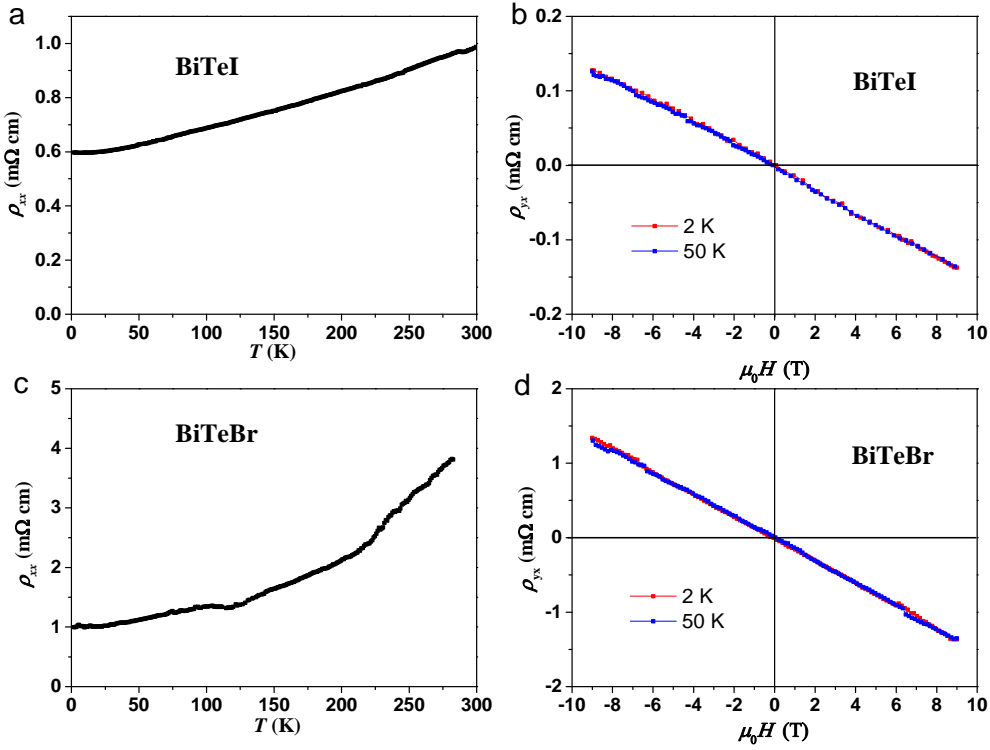

**Supplementary Figure 1. Resistivity and Hall resistivity of BiTeI and BiTeBr.** (**a**) and (**c**) show the temperature-dependent resistivities ($\rho_{xx}$) of BiTeI and BiTeBr, respectively. Both compounds display a metallic temperature dependence, the residual resistivity is higher for the BiTeBr sample than for BiTeI, suggesting a higher impurity concentration. (**b**) and (**d**) show the Hall resistivity, $\rho_{yx}(H)$ as function of magnetic field for BiTeI and BiTeBr, respectively. The $\rho_{yx}(H)$ curves have negative slopes at all temperatures, indicating that the conduction of both compounds is dominated by the electron-like charge carriers.

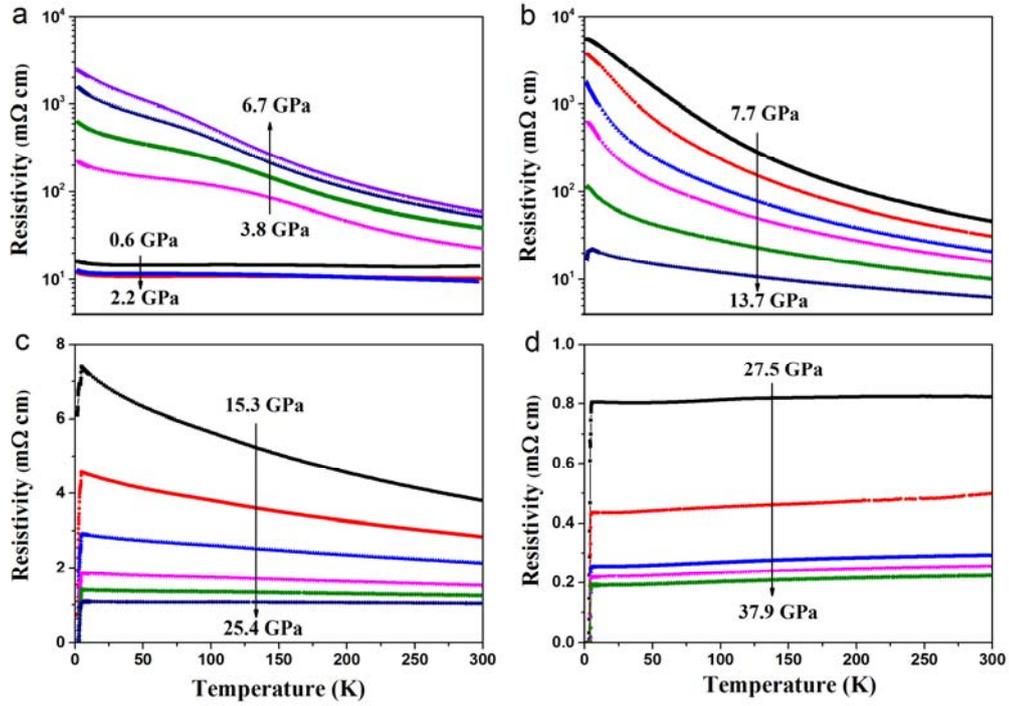

**Supplementary Figure 2. Evolution of the electrical resistivity as function of pressure for BiTeBr.** Similar behavior is observed for the BiTeBr system. Surprisingly, $\rho$ increases by almost three orders of magnitude and exhibits insulating behavior at pressures up to $\approx 8$ GPa[15]. As pressure increases to $\approx 14$ GPa, superconductivity is observed while the resistivity at higher temperatures still exhibits semiconducting behavior. Superconductivity persists at least up to pressures near 40 GPa and a maximum $T_c$ of 4.8 K is attained at $P = 31.7$ GPa.

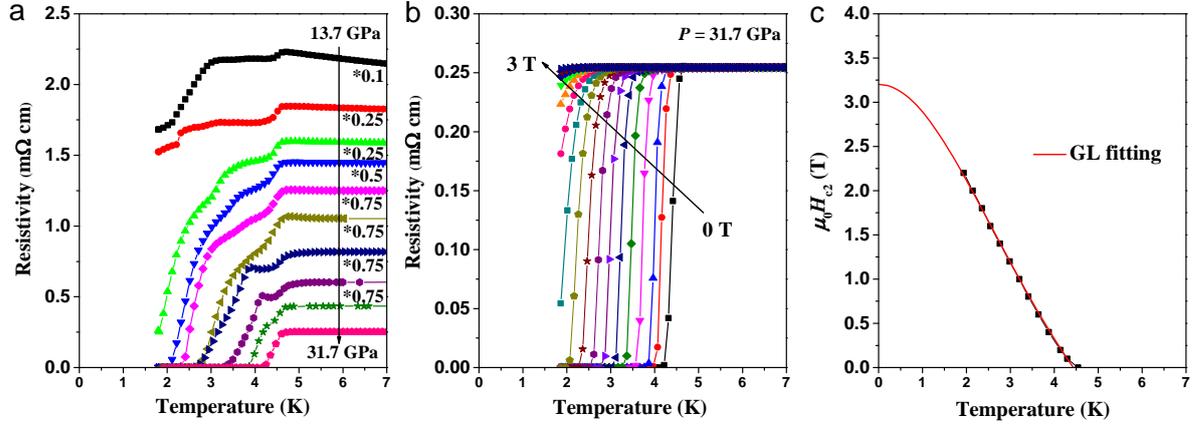

**Supplementary Figure 3. Temperature-dependent resistivity in the vicinity of the superconducting transitions and determination of the upper critical field for BiTeBr.** (**a**) A two-step transition is observed in the resistivity of BiTeBr, implying that two superconducting phases coexist in the sample. The $T_c$ of lower-$T_c$ phase (SC-I) increases gradually with increasing pressure. Only one sharp transition is observed for $P > 30$ GPa. We may conclude that phase SC-I is a bulk superconductor whereas phase SC-II (the one with the higher $T_c$ which does not display zero-resistance behavior) is not. There are a number of possibilities for the origin of phase SC-II. One is surface superconductivity[30, 31], another is a pressure-induced metastable phase with different crystal structure which coexists with phase SC-I. (**b**) Temperature dependence of the resistivity at different magnetic fields at 31.7 GPa. (**c**) Temperature dependence of the upper critical field (phase SC-I). Here, $T_c$ is determined as the 90% drop of the normal state resistivity. The solid lines represent fits based on the Ginzburg-Landau (GL) formula. $\mu_0 H_{c2}$ was extrapolated to 3.2 T, which yields a coherence length $\xi_{GL}(0)$ of ~10 nm.

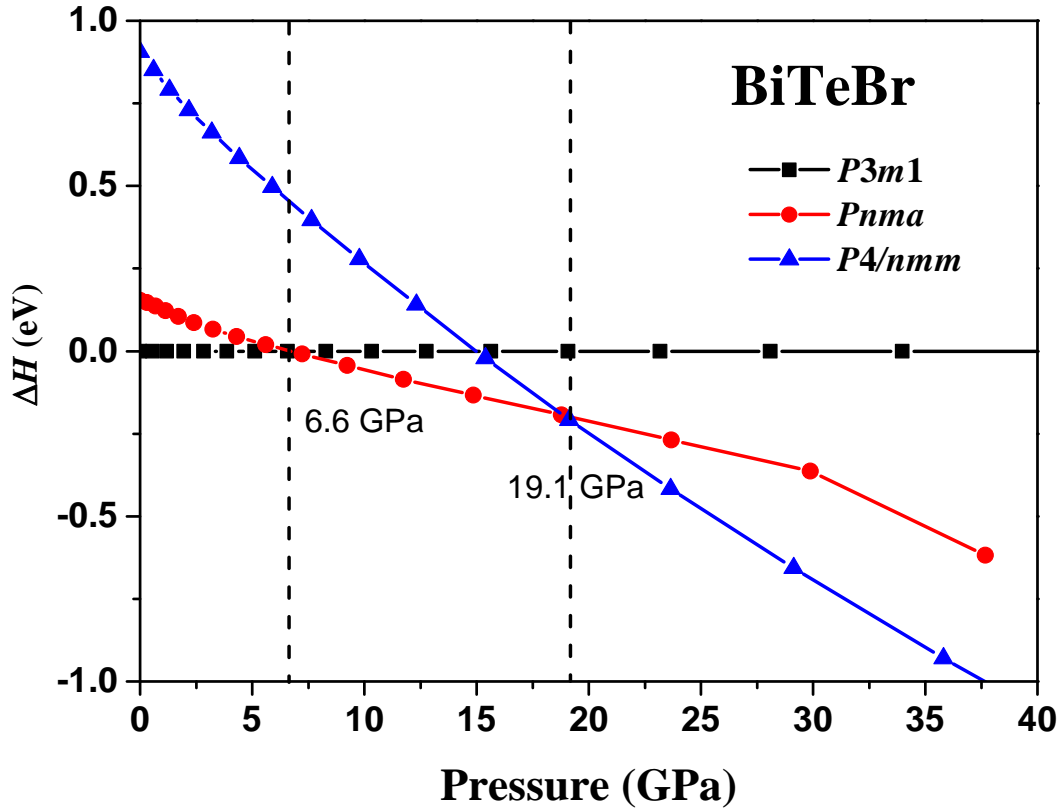

**Supplementary Figure 4. Enthalpy (relative to the *P*3*m*1 structure) of various structures as function of pressure for BiTeBr.** The enthalpies are given per formula unit. Enthalpy $H$ of a given phase was evaluated in order to identify the energetically favored ground state for a finite $P$ using $H = E_{tot} + PV$, where $E_{tot}$ is the total energy of the system and $V$ is the volume of the unit cell. The obtained $H$-$P$ curves indicate that the *P*3*m*1 structure is indeed the most stable at ambient pressure. In the $P$ range of approximately 6.6 to 19.1 GPa, the *Pnma* phase has the lowest enthalpy; however, for higher pressures, the *P*4/*nmm* phase is the ground state.

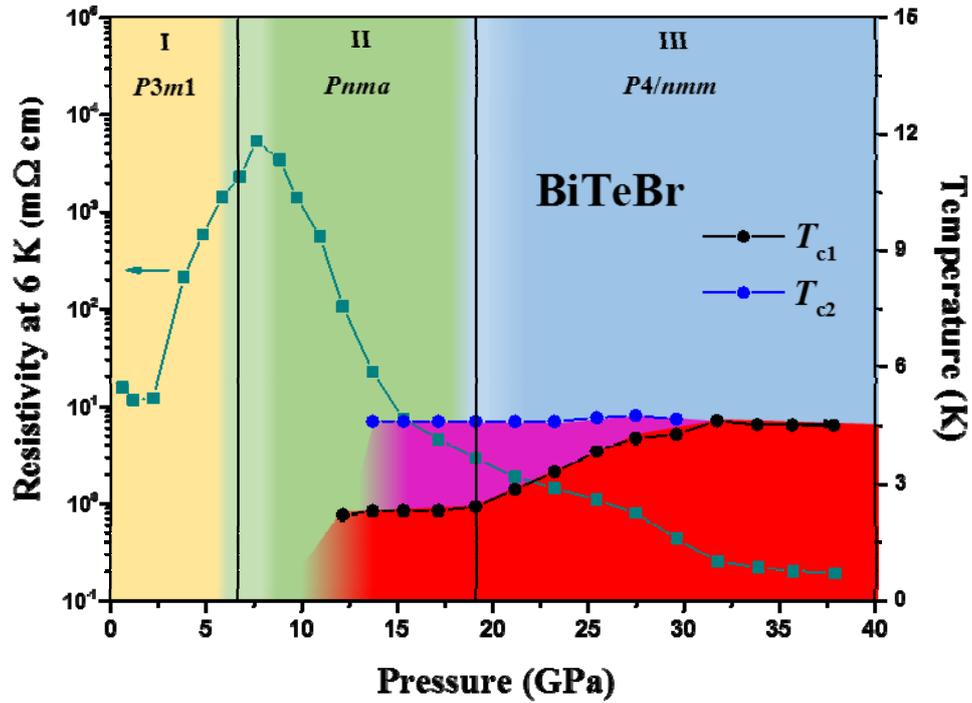

**Supplementary Figure 5. Electronic *P-T* phase diagram for BiTeBr.** A phase diagram similar to BiTeI is constructed for BiTeBr. A two-step transition is observed for BiTeBr. Red and pink areas are a guide to the eye indicating the superconducting phase SC-I and phase SC-II (cf. Supplementary Fig. 3).

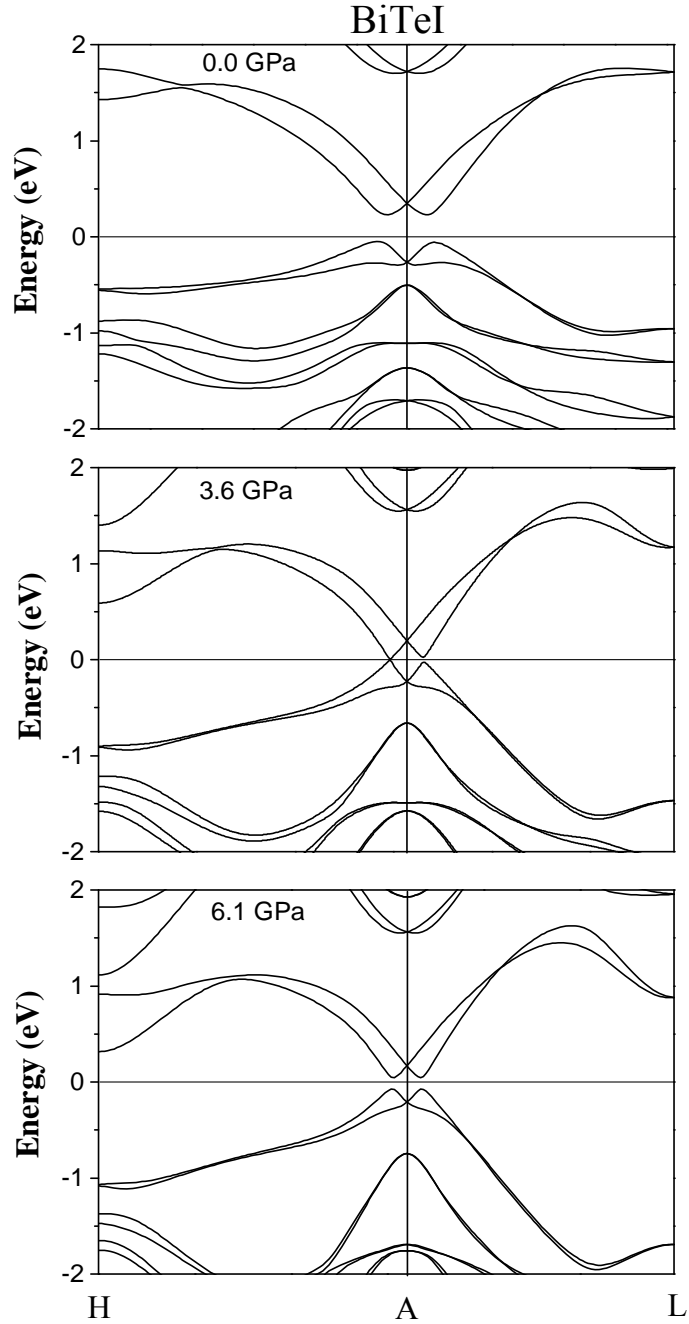

**Supplementary Figure 6. Electronic structures of phase I (*P*3*m*1 structure) of BiTeI at ambient pressure, 3.6 GPa and 6.1 GPa, respectively.** With increasing pressure the band gap decreases to zero until a critical pressure $P_c$ is reached. The band gap then reopens via inversion of the valence and conduction band characters, indicating a topological quantum phase transition at $P_c$ = 3.6 GPa for BiTeI.

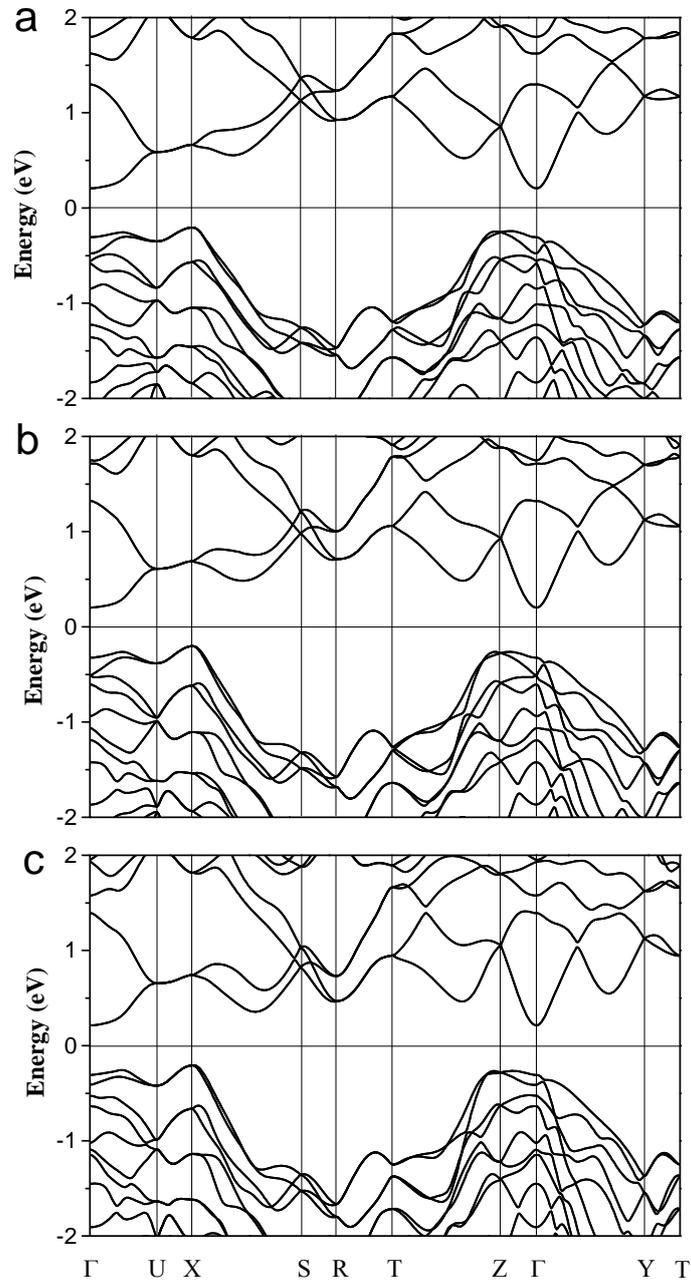

**Supplementary Figure 7. Calculated band structure in the *Pnma* phase of BiTeBr at 7.23 GPa (a), 11.74 GPa (b) and 18.78 GPa (c), respectively.** The conduction band at the R point of the Brillouin zone is sensitive to the pressure and shifts towards the Fermi level with increasing pressure.

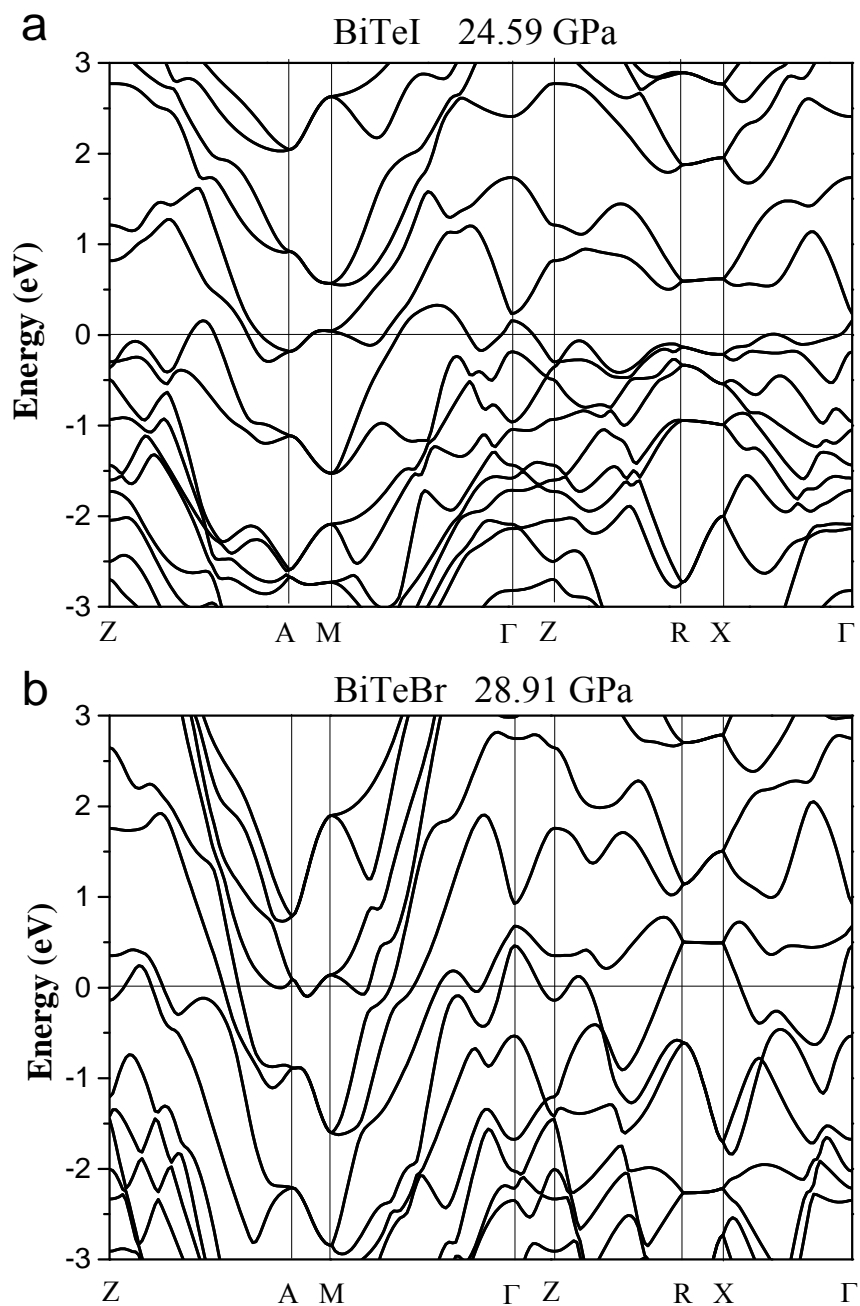

**Supplementary Figure 8. Electronic band structure of the *P*4/*nmm* phase of BiTeI and BiTeBr at 24.59 GPa and 28.91 GPa, respectively.**